\documentclass[,compatible]{spie}
\usepackage[utf8]{inputenc}
\usepackage{graphicx}

\makeatletter

\providecommand{\tabularnewline}{\\}

\usepackage{xcolor}

\newcommand{\bw}[1]{\iffalse{{#1}}\fi}

\@ifundefined{showcaptionsetup}{}{%
 \PassOptionsToPackage{caption=false}{subfig}}
\usepackage{subfig}
\makeatother

\begin{document}

\author{Andrew R. Willis\supit{a}, Md Sajjad Hossain\supit{a} and Jamie
Godwin\supit{b} \skiplinehalf \supit{a}University of North
Carolina at Charlotte, 9201 University City Blvd., Charlotte, NC~~28223
\\
 \supit{b}Air Force Research Laboratory, Munitions Directorate,
Eglin AFB, FL, 32542 \\
 }

\authorinfo{Further author information: (Send correspondence to A. Willis)\\
 A. Willis: E-mail: arwillis@uncc.edu, Telephone: 1 704 687 8420}

\title{Hardware-Accelerated SAR Simulation with NVIDIA-RTX Technology}
\maketitle
\begin{abstract}
Synthetic Aperture Radar (SAR) is a critical sensing technology that
is notably independent of the sensor-to-target distance and has numerous
cross-cutting applications, e.g., target recognition, mapping, surveillance,
oceanography, geology, forestry (biomass, deforestation), disaster
monitoring (volcano eruptions, oil spills, flooding), and infrastructure
tracking (urban growth, structure mapping). SAR uses a high-power
antenna to illuminate target locations with electromagnetic radiation,
e.g., 10GHz radio waves, and illuminated surface backscatter is sensed
by the antenna which is then used to generate images of structures.
Real SAR data is difficult and costly to produce and, for research,
lacks a reliable source ground truth. Few SAR software simulators
are available and even less are open source and can be validated.
This article proposes a open source SAR simulator to compute phase
histories for arbitrary 3D scenes using newly available ray-tracing
hardware made available commercially through the NVIDIA's RTX graphics
cards series. The OptiX GPU ray tracing library for NVIDIA GPUs is
used to calculate SAR phase histories at unprecedented computational
speeds. The simulation results are validated against existing SAR
simulation code for spotlight SAR illumination of point targets. The
computational performance of this approach provides orders of magnitude
speed increases over CPU simulation. An additional order of magnitude of GPU acceleration when simulations are run on RTX GPUs which include hardware specifically to accelerate OptiX ray tracing. The article
describes the OptiX simulator structure, processing framework and
calculations that afford execution on massively parallel GPU computation
device. The shortcoming of the OptiX library's restriction to single
precision float representation is discussed and modifications of sensitive
calculations are proposed to reduce truncation error thereby increasing
the simulation accuracy under this constraint.
\end{abstract}


\section{Introduction}

\bw{What is SAR, Why is it important}Synthetic Aperture Radar (SAR)
is an image formation process that uses pulses of electromagnetic
radiation emitted and sensed by a high-power antenna from a moving
platform \cite{radartutorial}. Application of SAR includes target
recognition \cite{TargetRecognition}, mapping difficult terrain area
\cite{Mapping}, ocean surveillance \cite{Ocean_Surveillance}, geology
(structural mapping, lithological mapping, mapping geomorphological
features, mineral exploration, active fault mapping) \cite{Geology},
mapping forest cover and biomass \cite{ForestMapping}. %\aw{ADD - Applications of SAR and citations}

\bw{?? Intro paragraph}SAR imaging uses small antenna to illuminate
large swaths of ground from different antenna locations \cite{mitocwSAR}.
The transmitted electromagnetic wave propagates and interacts with
matter or objects. Optical sensors, using visible light, can only
function in the daytime and measure reflected solar light. These sensors
lose their utility when the surveyed region is obstructed by weather
such as cloud coverage. Fortunately, microwaves can operate in day
or night in nearly all weather conditions because of its ability to penetrate through clouds and vegetation (depending on frequency).
Penetration through the forest canopy or into the soil is greater
with longer wavelengths. It also has minimal atmospheric effects with
minimal sensitivity to structure and dielectric properties. The received
reflected data can be processed into a ground image using various
SAR signal processing algorithms \cite{Carrara1995}. Each pixel of
a radar image represents a complex quantity of energy that was reflected
back to the antenna. The magnitude of each pixel represents the intensity
of the reflected signal.

\bw{?? Intro paragraph2}There are substantial challenges associated
with SAR simulation. The critical components of SAR sensing are vehicle
pose/trajectory, antenna aperture, SAR signal processing system, electromagnetic
interaction with matter, and scene geometry. The scene has to be constructed
which often uses a polygonal representation (triangle mesh) that define
the shape and reflectance properties. The antenna must be defined
next to establish its placement within the \char`\"{}scene\char`\"{}
and must replicate signal responses that coincide with the region
of the scene that would be observed by the antenna. The simulated
EM waves must propagate into the scene and calculate how they refract,
attenuate, and reflect off scene surfaces. The computational complexity
of the problem grows significantly as scenes become more intricate,
having millions of polygons to trace, and the number of EM wave bounces
that wish to be preserved.

There are multiple high-investment barriers that make generation of
real-world SAR data difficult. SAR antenna typically emit high power
pulse signals that exceed FCC broadcast regulations and as such require
licensing to operate. SAR pulse emission and echo sensing typically
require unobstructed views for distances $>$ 5 km. This requirement
restricts deployment to aerial platforms. Further, signal processing
algorithms for SAR require high-precision and high data rate knowledge
of the vehicle position and orientation which requires costly positioning
and guidance, navigation and control software. The radio frequency
hardware (RF signal conditioning and antenna) is often extremely expensive, 
contains sensitive and highly-specialized circuitry, computing hardware
and RF antenna hardware. Hence, cost, licensure, equipment, personnel,
training and expertise makes the total cost-to-own for SAR technology
prohibitive in reality. All of these barriers for SAR signal generation
motivate the creation of software systems that can simulate the SAR
sensing context.

\bw{Why SAR simulation?}Collected SAR data is subject to numerous
noise source including EM noice, position error, non-planar targets,
and range curvature. These errors make ground truth difficult to establish
and hinder theoretical validation of SAR models for EM propagation
and backscatter computation. Availability of simulated data with ground
truth or deliberately introduced noise allows theoretical validation
of SAR signal process and image reconstruction
algorithms. For this reason, simulation data is often the preferred choice
in research and development communities due to its versatility and
smaller cost relative to collected measured data. Simulation tools
provide the opportunity to mimic the effects of any radar platform
in any environment under any specified conditions so long as the simulator
has been developed to respond to such conditions.

\bw{applications}Our SAR simulation goals seek to support two applications.
The first application uses synthetically generated SAR images to facilitate
deep learning SAR research and bolster the machine learning community
with the capacity to incorporate synthetic data methods \cite{2015arXivHowMuchData,Tremblay_2018_CVPR_Workshops}.
The second application is to provide high speed SAR simulation of
raw radar signals for real-time simulation of SAR imagery.

\bw{contribution} This work transforms and extends current electromagnetic
(EM) wave propagation simulation models for SAR image formation using
a low-cost, massively parallel and hardware accelerated computing
architecture provided by NVIDIA’s recent RTX series of GPUs \cite{AnderssonNSSA19}
in an effort to accelerate SAR simulation
capabilities.

\section{Related Work}

\bw{Overview}SAR simulation requires solutions to far-field Electro-Magnetic
(EM) wave propagation equations. Fortunately, these equations can be
solved for approximately by breaking the complete solution into two
parts: (1) solving for the geometric propagation of EM waves through
3D space, termed geometrical optics, and (2) solving for the physical
interactions between EM waves and scene surfaces, termed physical
optics. Our review of current state-of-the-art discusses the dominant
simulation theoretical model, current implementations of this model
and how this article advances state-of-the-art in this domain.

\bw{SBR approaches}From a theoretical standpoint, current approaches
overwhelmingly adopt the Shooting and Bouncing Ray (SBR) method \cite{SBR_Lee}.
These methods use geometric optics to simulate
surface scattering of incident EM waves. Geometric optics theory use linear
rays to model the direction of wave propagation and attributes each ray with a packet of energy distributed within a cylindrical
region about a 3D ray, referred to as a \emph{ray tube} \cite{RayTube_Lee}.
Ray tube simulation is the basis of all SAR simulation techniques
described in this article and is the dominant computational model
for calculating EM wave scattering from surfaces for the purposes
of simulating SAR wave propagation and, more generally, the Radar
Cross Section (RCS) of scattering surfaces.

\bw{Commercial approaches}From a technical standpoint, current SAR
simulation implementations are very limited consisting of few commercial
\cite{Andersh2000Xpatch4T,OSV_Radar,Raytheon} and even less open-source
\cite{StephenAuerPhDthesis,Weijie2014SARIS} implementations. Commercial
implementations pose difficult problems for validation, improvement
and extension and may also be unavailable for public use (governmental
or military use) or prohibitive in cost. Open-source approaches have
limited application as they do not solve major fundamental technical
technical problems including the ray tracing component of simulation
\cite{Weijie2014SARIS}.

\bw{XPatch}One example of a commercialized SAR simulator described
in the literature is Xpatch \cite{Andersh2000Xpatch4T}. Xpatch is
a validated radar simulation software package used by industry and
government to simulate SAR images. As mentioned previously, the Xpatch
simulator uses 3D models and simulates EM waves using the Shooting
and Bouncing Ray (SBR) method. Xpatch provides the user the ability
to generate Radar Cross Section (RCS), High Resolution Range (HRR)
profiles, and Synthetic Aperture Radar (SAR) imagery. Unfortunately,
the tool is proprietary, therefore does not lend the option for open-source
use.

\bw{RaySAR}The leading example of an open-source SAR simulator is
RaySAR \cite{RaySAR3D}. RaySAR uses custom modifications to the open
source POV-Ray (Persistence Of Vision) ray tracing engine to model
EM wave propagation, reflection and backscatter \cite{alma991010725391504091,StephenAuerPhDthesis}.
RaySAR requires the user to employ the RaySAR-customized POV-Ray software
to solve for geometric EM wave and surface intersection locations
that track wave intensity and direction as the EM waves bounce across
potentially multiple surface-to-surface reflection paths resulting
in a (sometimes large) output data file. RaySAR's main interface uses
MATLAB to analyze the POV-Ray generated data file and compute the final
simulation result. Shortcomings of this approach are: (1) MATLAB is
required to run the simulator, (2) POV-Ray uses CPU ray tracing which
can be slow and also generate very large data files for complex 3D
scenes and (3) the simulator only uses ray traced solutions for a single
view which makes the simulated reconstruction inaccurate; especially
when the aperture is large.

\bw{contribution}In contrast to these approaches, we propose a completely
new computational paradigm for solving the geometrical optics problem
of SAR simulation.
\begin{enumerate}
\item Our proposed OptiX SAR simulator is open source and does not require
third party programs, e.g., MATLAB. 
\item Our proposed OptiX SAR simulator uses extremely fast GPU parallel
processing to compute the ray traced solution. 
\item Our proposed OptiX SAR simulator calculates ray traced solutions for
multiple views making the simulated reconstruction accurate even when
the aperture is large.
\end{enumerate}
The cumulative impact of these contributions is a new SAR simulation
framework that applies to most airborne SAR sensing contexts that
can simulate SAR backscatter from complex 3D scenes at unprecedented
performance rates. 

\section{Background}

\bw{Overview, what is in the background, why it is a required in
this article} Our approach makes extensive use of the OptiX ray-tracing
library to achieve unprecedented performance in calculating the geometric optical propagation for SAR simulation. In this section
we provide a brief overview of the OptiX system and detail its unique
massively-parallel execution structure. An understanding of this structure
is crucial to the development of our proposed OptiX-based SAR simulation
approach.

\subsection{OptiX Ray-Tracing Engine}

\bw{OptiX, what is it, why is it useful} In this paper, we will
utilize the OptiX library \cite{NvidiaRTXdev} from NVIDIA and special-purpose
NVIDIA RTX GPU hardware to provide hardware-acceleration and enhancing
computational speed for calculating ray tracing results for scenes
having arbitrary 3D structure. RTX technology has been shown to successfully
accelerate similar applications in computer graphics where deployment
of RTX GPU-hardware acceleration \cite{AnderssonNSSA19} offers an
increase of approximately $3.5X$ beyond non-RTX technology for real-time
ray tracing.

\bw{OptiX program structure} OptiX \cite{NVIDIA_OptiX_6} ray tracing
engine has seven different types of programs, i.e, ray generation,
intersection, closest hit, any hit, miss, exception and selector visit.
In addition, a bounding box program operates on geometry to determine
primitive bounds for accelerated structure construction. The core
operation, rtTrace, alternates between locating an intersection (Traverse)
and responding to that intersection (Shade). GPU support includes geometry
acceleration and scenegraph (which can be dynamically modified). Also,
GPU acceleration for custom intersection processing (shaders) includes
a Miss program (rays that miss targets), Closest hit program (reflection,
surface interactions) and Any hit program (shadows).

\begin{figure}[htp!]
\noindent \begin{centering}
\subfloat[Algorithm]{\noindent \begin{centering}
\includegraphics[height=2in]{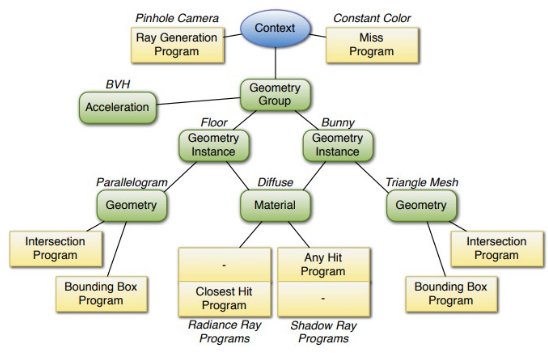} 
\par\end{centering}
}\qquad{}\subfloat[Illustrated example]{\noindent \begin{centering}
\includegraphics[height=2in]{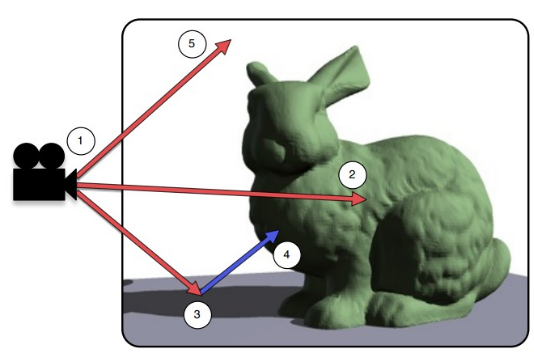} 
\par\end{centering}
}
\par\end{centering}
\caption{OptiX ray tracing of an example scene with a pinhole camera, two objects
and shadows (a) algorithm (b) illustrated example (taken from \cite{Parker_optix:a}).}
\label{fig:OptiX ray tracing of an example scene} 
\end{figure}

\bw{An example of OptiX at work} Figure \ref{fig:OptiX ray tracing of an example scene}
from S. Parker et al. \cite{Parker_optix:a} depicts the operating
principles of massively parallel OptiX raytracing. \ref{fig:OptiX ray tracing of an example scene}(a)
shows a hierarchical organization of objects that makeup the OptiX
context. Of particular note are the geometry instances and material
instances. Geometry instances denote collections of polygons (typically
triangle meshes) which will interact with traced rays. The specific
interactions of the geometry instances depends upon the materials
attributed to the geometry, e.g., green diffuse bunny and grey ground
plane. The OptiX engine decomposes geometry into local sub-groups
using a Bounding Volume Hierarchy (BVH) \cite{BVH} to accelerate
the search for ray-geometry intersection locations. 

Figure \ref{fig:OptiX ray tracing of an example scene}(b) shows how rays
are traced into a 3D scene by the OptiX engine. It depicts the five
categories of rays that OptiX uses. To allow massively parallel raytracing,
OptiX allows users to design small programs which can be associated with
each ray category which are invoked as required to generate a raytracing
result. Provided the example in Figure \ref{fig:OptiX ray tracing of an example scene}, the OptiX engine operates in the following fashion:
\begin{enumerate}
\item A \emph{ray generation} program is a starting point for execution.
This program creates and traces rays into the 3D scene. Each ray initiates
a BVH traversal to determine how the ray interacts with geometries
of the context. 
\item A \emph{closest hit} program will be called for those rays that intersect
scene geometries. OptiX invokes the program and passes it the 3D intersection location and the local surface geometry which can be used to modify
the ray tracing result or emit new rays. 
\item A \emph{any hit} program will be called when a ray intersects the
scene multiple times which is useful for computation of surface shadows
and surfaces having transmissive material properties, e.g., glass
simulation. 
\item A \emph{miss} program will be called for those rays which do not intersect
any scene geometries. 
\item An \emph{exception} program will be called when unexpected ray tracing
circumstances are detected by the OptiX engine. 
\end{enumerate}
The OptiX library integrates these programs to a structure purposefully
designed to minimize dependency between traced rays which, by design,
produce high performance ray tracing results for potentially complex
phenomenon. Performance gains are reaped via massively parallel computation
of programs that run asynchronously and within limited memory to perform
local atomic contributions to the final result.

\section{Methodology/ Research Description}

Our implementation of SAR simulation adopts the NVIDIA OptiX library
to compute the solution to the geometric optics problem. We break
the complete simulation problem into (5) main parts: 
\begin{enumerate}
\item Load the virtual 3D world into a OptiX context and send it to the
GPU compute device(s). 
\item Initialize models of the vehicle trajectory, pulse signal waveform
and radiation characteristics of the SAR antenna. 
\item Simulate the emission of SAR EM pulse and sensed backscatter by the
antenna. 
\item Advance the vehicle to the next position within the trajectory.
\item Return to step (3) until the backscatter from all pulses has been
calculated; then exit.
\end{enumerate}
The models applied in this article rephrase existing SAR simulation
models to enable massively parallel and hardware accelerated computation
of the backscatter signal through the OptiX library and its hardware
realization on NVIDIA RTX GPUs.

\subsection{OptiX Simulation}

The SAR simulation software contains two parts; simulate the scene
and simulate a SAR system. A 3D scene must be generated and loaded
onto the GPU device(s). The scene contains all the models used to
generate a simulated environment. This comprises of all the objects,
including the basic material properties of each object that the user
wishes to generate a radar response from. The scene also contains
a description of the radar system which characterizes the antenna
and radar platform. An observation view is also established to provide
the user an overall view of the entire simulation, allowing the user
to simultaneously monitor the radar and scene simulations as desired.
The most complex portion is the SAR simulation which outlines how
the phase history will be simulated using ray tracing to complete
the SAR simulation. An in-depth description of the OptiX simulation
is provided in the following sections.

\subsection{Creating a Virtual 3D World into a OptiX context}

We define 3D scenes using YAML format files \cite{YAML}. Our simulator
parses the simulation YAML file to virtually construct the 3D scene.
Virtual worlds for SAR simulation include at least one instance of
each of the following (5) objects: 
\begin{itemize}
\item One (or more) descriptions of 3D scene geometries. 
\item One (or more) descriptions of scene lights. 
\item One set of SAR antenna parameters. 
\item One set of SAR signal processing system parameters. 
\item One set of vehicle trajectory and aperture parameters. 
\end{itemize}
There are a total of (8) different YAML objects for the simulator.
Scene geometries and their reflectance are specified using a combination
of (3) YAML objects. The \char`\"{}mesh\char`\"{} and \char`\"{}primitive\char`\"{}
YAML objects specify object geometry and \char`\"{}material\char`\"{}
YAML objects specify reflectance properties for geometries. The \char`\"{}light\char`\"{}
and \char`\"{}camera\char`\"{} YAML objects serve to illuminate and
visualize the virtual 3D scene (in the observation window) respectively.
Finally the \char`\"{}antenna,\char`\"{} \char`\"{}sarsystem,\char`\"{}
and \char`\"{}trajectory\char`\"{} YAML objects specify the antenna,
signal processing system and vehicle trajectory and aperture parameters
respectively. Since YAML files are plain text, users can easily modify
and recombine simulation configurations to create complex SAR simulations
of 3D scenes. 
\begin{table}
\noindent \begin{centering}
\begin{tabular}{|c|l|}
\hline 
\textbf{YAML Object}  & \textbf{Properties} \tabularnewline
\hline 
\hline 
camera  & pose, $(x,y)$ resolution (pixels), $(x,y)$ field of view (degrees)\tabularnewline
 & and projection type: \{pinhole, orthographic\}\tabularnewline
\hline 
primitive  & pose, material, size and shape type: \{plane, box, sphere, torus\}\tabularnewline
\hline 
mesh  & pose, material, size, filename and file format: \{OBJ, PLY\}\tabularnewline
\hline 
material  & shader program: \{diffuse, specular, specular-transmissive, phong\},\tabularnewline
 & and reflectance parameters\tabularnewline
\hline 
light  & pose, size, intensity, and light color\tabularnewline
\hline 
antenna  & pose, intensity, size, shape, efficiency and type: \{transceiver,
transmitter, receiver\}\tabularnewline
\hline 
sarsystem  & carrier frequency (GHz), range resolution (m), range swath (m),\tabularnewline
 & number of frequencies, polarity type: \{HH, HV, VH, VV\}\tabularnewline
\hline 
trajectory  & pose (as slant range (m) at point of closest approach and depression
angle (deg))\tabularnewline
 & $(x,y,z)$ target position, Synthetic aperture size as Azimuth start
and end (deg)\tabularnewline
 & Number of pulses, and SAR mode: \{Linear Spotlight, Circular Spotlight,
Stripmap\}\tabularnewline
\hline 
\end{tabular}
\par\end{centering}
\caption{\label{tab:YAML_Objects}Simulation text file YAML objects and their
parameters.}
\end{table}

Each YAML object has a variety of options. Table \ref{tab:YAML_Objects}
summarizes the properties of the YAML objects. The SAR antenna is
specified by describing the operating bandwidth and the mode i.e.
spotlight and stripmap. If we consider a scenario where the scene
is stationary, the antenna has to traverse to generate simulated phase
history data used for SAR image formation. 

\subsubsection{Describing and Visualizing the Virtual 3D Scene}

Virtual 3D scenes are constructed by adding 3D geometries to the scene
and attributing these geometries with reflectance attributes to determine
their radar cross section. Each YAML file also includes camera and
light objects to allow the 3D scene structure to be visualized using
conventional ray tracing. The user can navigate the scene in this window
to view and visually validate the structure of the simulated 3D scene.

\begin{figure}
\noindent \begin{centering}
\subfloat[pixel samping]{\begin{centering}
\includegraphics[height=1.5in]{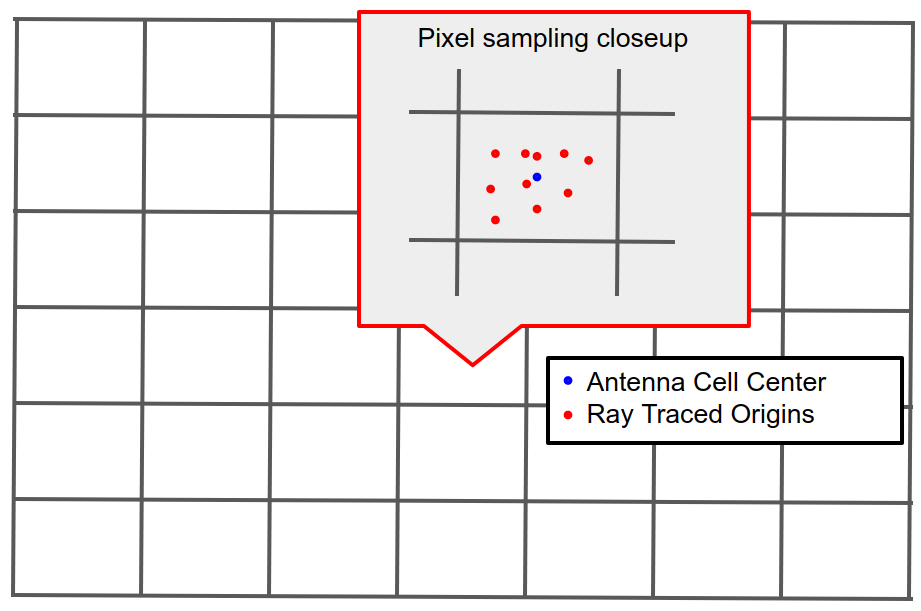} 
\par\end{centering}
}\qquad{}\subfloat[A visualization of a SAR virtual world.]{\begin{centering}
\includegraphics[height=1.5in]{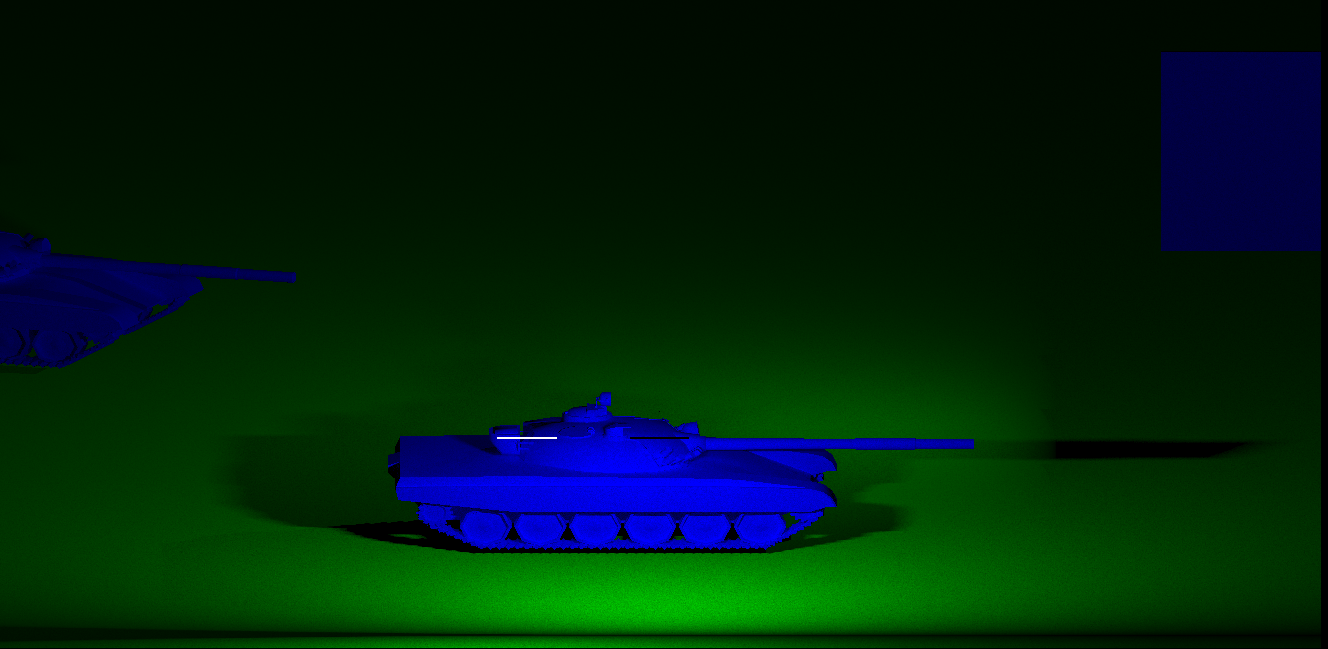} 
\par\end{centering}
}
\par\end{centering}
\caption{\label{fig:YAMLObjects}(a) shows details of pixel image formation
via sampling the scene with perturbed ray origin positions within
a pixel. (b) shows a visualization of a simple scene using these image
formation principles.}
\end{figure}

A summary of the capabilities of the simulator for 3D scene creation
and visualization are listed below:
\begin{itemize}
\item Camera - sets the observer view when the scene is shown in the interactive
window 
\item Lights - lights required to view the scene as an observer 
\item Objects 
\begin{itemize}
\item Primitives - spheres, boxes, planes, torus and parallelogram 
\item Mesh - we can load and place 3D mesh models into the 3D scene. 
\item Materials - objects are attributed with materials to model how they
reflect incident light and SAR radiation. 
\end{itemize}
\end{itemize}
\bw{3D Scene Visualization}Our ray tracing method for visualization
(and SAR simulation of) 3D scenes seek to compute pixel color and
intensity by calculating the integral of the radiated energy incident
to the surface of each pixel. To do so, we trace rays from each pixel
into the 3D scene and, when intersections are found, we determine
the amount of energy reflected onto each pixel in the (Red, Green,
and Blue) spectrum. Calculation of the integral is computationally
costly so an approximation that is both lower cost and compatible
with parallel processing is provided by using Monte-Carlo/Metropolis
integration. As shown in Figure \ref{fig:YAMLObjects}(a), we perturb the
origin of each traced ray which results in more accurate renderings
of the simulated image. This technique is particularly suited for
photon counting, i.e., energy tracking, rendering methods often used
for photo-realistic ray tracing as in McGuire et al. \cite{McGuire2009Photon}.
Figure \ref{fig:YAMLObjects}(b) shows a scene visualization using
this approach that includes several 3D mesh objects: 2 tanks having
blue diffuse reflectance, 1 box primitive also having blue diffuse
reflectance, and 1 plane primitive having a green diffuse reflectance
(the ground plane/background).

\subsubsection{Describing the SAR Emission and Sensing Process}

A description of the virtual SAR sensing context is comprised of 3
parts: (1) a description of the SAR antenna, (2) a description of
the SAR signal processing system and (3) a description of the vehicle
trajectory.

The SAR antenna is considered to be mounted on the vehicle having
the shape and dimensions specified by the antenna YAML object. We
model the physical aperture of the antenna similar to that of the
conventional camera as described in Figure \ref{fig:YAMLObjects}(b).
Yet, the field of view and resolution of the antenna cells are determined
by the antenna radiation pattern. Our simulator assumes radiated energy
out of the antenna aperture is sufficiently large only over the main
lobe of the antenna radiation pattern. Hence, the field of view is
taken as the angular span between the first pattern nulls adjacent
to the main lobe, referred to as the ``First Null Beam Width.''
Simulation currently supports antenna or antenna arrays that can be
modeled as a single rectangular patch. Depending upon the SAR measurement
mode, the antenna will either remain static (stripmap SAR) or actuate
as the vehicle moves (scanning, spotlight SAR). The antenna intensity
parameter is used to distribute radiated energy across the aperture
creating a unique energy for each aperture cell.

The SAR signal processing system characterizes the control timing
and structural content of the signals radiated by the SAR antenna.
This includes the polarity of the emitted and sensed radiation, each
of which can be Horizontal (H) or Vertical (V) with respect to the
antenna orientation generating four polarimetric modes: \{HH, HV,
VH and VV\}. If nothing is specified, it takes the default value (sum
all polarization states). Additional signal processing parameters
specify the carrier frequency or operating frequency in GHz, the bandwidth
of the emitted radiation specified as the desired range resolution
and the discrete resolution of the sensed signal as the number of
frequencies sensed in the return signal. The range resolution determines
the bandwidth, i.e., range of frequencies emitted in each radiated
SAR chirp pulse and determines the range resolution of the SAR system
by descretizing the range swath into range cells as shown in Figure
\ref{fig:SAR-trajectory-and-signals}(b).

The vehicle trajectory is determined as a combination of the SAR mode
and geometric positioning information. This information determines
a collection of discrete locations from which the vehicle antenna
emits pulses onto the target. The simulator allows four different
SAR modes:
\begin{enumerate}
\item Spotlight (linear flight path with antenna and focal point movement), 
\item Circular Spotlight (circular flight path with no antenna and focal
point movement), 
\item Scanning (linear flight path with antenna and focal point movement),
and 
\item Stripmap (linear flight path with no antenna and focal point movement). 
\end{enumerate}
Since all SAR modes have a common point of closest approach, this
$(x,y,z)$ position determines the vehicle's position at its midpoint
in the aperture. This is taken as the unique $(x,y,z)$ point on the
upper $(+z)$ hemisphere centered on the target point having radius
$\rho_{s}$, azimuthal angle $\theta$, and depression angle $\psi$.
The $(x,y,z)$ position is computed as a point in a spherical coordinate
system having coordinates $(\rho_{s},\theta,\phi)$ where $\phi$
is the look angle, i.e., $\phi=\frac{\pi}{2}-\psi$. The azimuthal
angle of closest approach is taken as the average of the aperture
azimuth start and end angle, $(\theta_{s},\theta_{e})$, i.e., $\theta=\frac{\theta_{s}+\theta_{e}}{2}$.
These parameters then jointly define a continuous model for the SAR
vehicle trajectory. The number of pulses parameter, $N_{p}$, determines
the sampling of this trajectory at discrete locations where pulses
are spatially distributed over the aperture trajectory at equal intervals
of arc-length for both linear or circular trajectories.

\subsection{SAR Simulation}

The second component to SAR simulation applies a model for EM wave
propagation to calculate the sensed antenna signals for each pulse
emitted from the antenna. SAR simulation is accomplished in the following
sequence of steps: 
\begin{enumerate}
\item Initialize the GPU device for SAR simulation and place the antenna
at the required pose at the aperture start.
\item SAR pulse energy is distributed across a $N\times M$ grid of cells
on the antenna aperture and OptiX ray tubes of descretized RF energy
are traced into the 3D scene.
\item OptiX computes scene intersection locations and the ``closest hit''
program computes their contribution to SAR phase history for that
pulse. 
\item The vehicle is moved to the next location in the trajectory and returns
to step (2) until the aperture end position is reached.
\end{enumerate}
Once complete, the simulated backscatter sensed from RF energy reflected
by scene surfaces is captured in the SAR phase history. This is the
raw signal used by downstream reconstruction algorithms to calculate
2D SAR images of scene objects.

\subsubsection{Intialization}

The SAR simulation is initialized by attributing all scene objects
with material properties that characterize their reflectance behavior
for the chosen SAR frequency, e.g., X-band SAR uses 10GHz EM radiation.
To preserve GPU device space and slightly improve performance, the
observation window is made inactive and the lights of the virtual
scene used for visualization are removed, e.g., ``turned off,''
to simplify the simulation.

A virtual antenna is created using the camera pixel model of Figure
\ref{fig:YAMLObjects}(a) to simulate the antenna aperture. Typically,
for modestly large aperture antennae this produces a virtual camera
view with a very long focal length due to the small field of view
or, equivalently, small divergence, of the antenna beam intensity.
The number of vertical pixels, or equivalently, range cells, is determined
by the number of frequencies (range resolution) from the simulator
settings. The antenna is placed such that it's phase center is located
at the vehicle trajectory starting position and oriented such that
the plane of the patch antenna is perpendicular to the vector from
the target position to the current vehicle position.
\begin{figure}
\noindent \begin{centering}
\subfloat[SAR vehicle trajectory]{\begin{centering}
\includegraphics[height=1.4in]{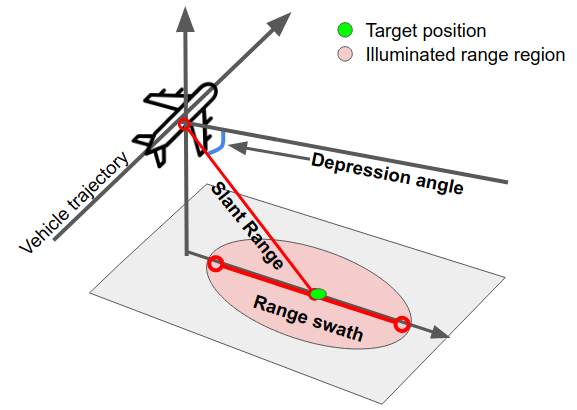} 
\par\end{centering}
}\subfloat[SAR pulse and range resolution]{\begin{centering}
\includegraphics[height=1.4in]{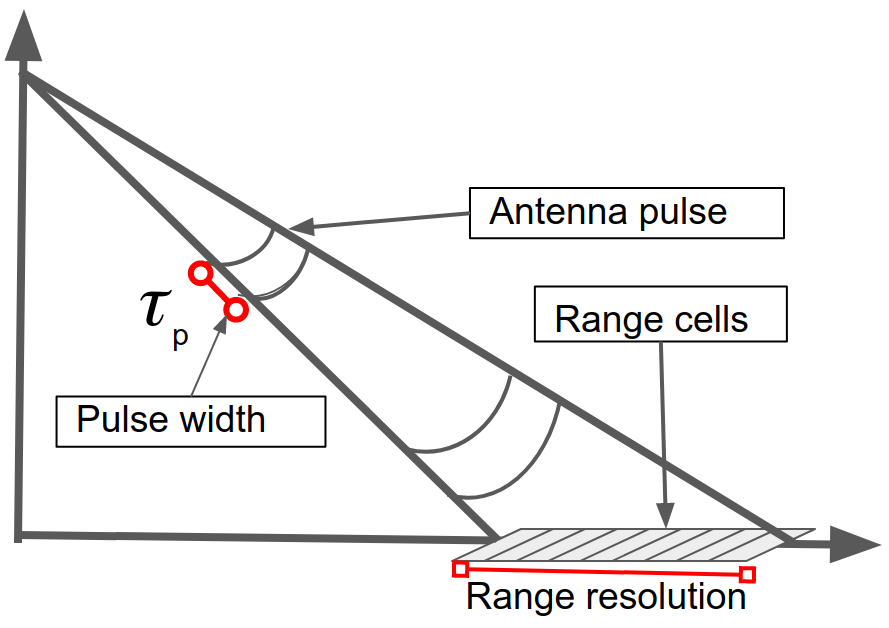} 
\par\end{centering}
}\subfloat[SAR differential range]{\begin{centering}
\includegraphics[height=1.4in]{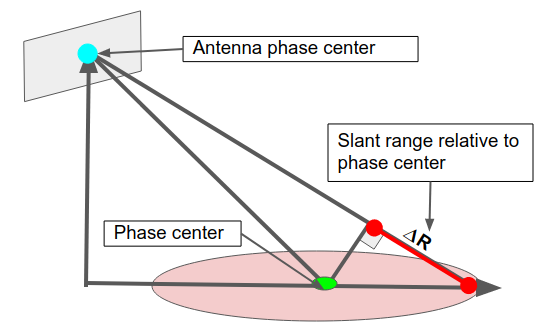} 
\par\end{centering}
}
\par\end{centering}
\caption{\label{fig:SAR-trajectory-and-signals}(a) shows the geometry of the
SAR vehicle trajectory. (b) shows the parameters of the SAR signal
pulses.}
\end{figure}

\subsubsection{Theoretical Model: SAR Pulse and Phase History Simulation}

We use the theoretical model of the Shooting and Bouncing Rays (SBR)
method to calculate the backscatter sensed by the antenna in response
to SAR pulses. The raw signal sensed by the antenna in response to
a pulse is referred to as an echo. Each echo contains geometric and
material information from the scene in the form of the scene backscatter
encoded in space by the time-of-arrival of EM frequencies through
observed magnitudes and phases of the emitted frequencies. Often this
echo is represented by its Fourier transform, $S(f_{k})$. For the
proposed OptiX simulator, there are $N_{p}$ pulses emitted from
the vehicle as it traverses the synthetic aperture. Let $\tau_{i}$
denote the time at which the $i^{th}$ pulse is emitted with $i=1,2,\ldots N_{p}$.
Using this notation we can construct a complex-valued 2D image $S(f_{k},\tau_{i})$
which is also referred to as the phase history. As such, the SAR simulation
problem reduces to the problem of computing the phase history, $S(f_{k},\tau_{i})$.

To compute the phase history of a single pulse we must compute EM
backscatter from the emitted antenna pulse. Assuming linearity, the
total reflected energy is the sum of reflected energy from all illuminated
surface positions. Note that, for planar surfaces, surface locations
of constant (slant) range lie on hyperbolic curves and direct, i.e.,
round-trip antenna-surface-antenna, backscatter will arrive contemporaneously
to the antenna and superimpose. To simplify the model, we eliminate
the instantaneous position of the vehicle from the model by considering
only the time-of-arrival relative to the target phase center, i.e.,
the middle of the range swath. 

Let $(x,y,z)$ denote the 3D position of the antenna phase center
and $p_{c}(\tau)=\left(x_{c}(\tau),\,y_{c}(\tau),\,z_{c}(\tau)\right){}^{T}$
denote the position of the target phase center at time $\tau$. Then
slant range between the antenna phase center and target location is
given by Equation (\ref{eq:range-to-swath-center}).
\begin{equation}
d_{c}(\tau)=\sqrt{(x_{c}(\tau)-x)^{2}+(y_{c}(\tau)-y)^{2}+(z_{c}(\tau)-z)^{2}}\label{eq:range-to-swath-center}
\end{equation}
Let $p_{t}(\tau)=\left(x_{t}(\tau),\,y_{t}(\tau),\,z_{t}(\tau)\right){}^{T}$
denote an arbitrary third scene point, referred to as a \emph{target}
\emph{location}, illuminated by SAR radiation. Similarly, the slant
range between the antenna phase center and this target location is
given by Equation (\ref{eq:range-to-target-location}).

\begin{equation}
d_{t}(\tau)=\sqrt{(x_{t}(\tau)-x)^{2}+(y_{t}(\tau)-y)^{2}+(z_{t}(\tau)-z)^{2}}\label{eq:range-to-target-location}
\end{equation}
The differential range, $\Delta R(\tau)$, captures the difference
in range (and sensed time of arrival at the antenna) between the target
scene point and the phase center scene point as shown in Equation
(\ref{eq:differential-range}) and depicted in Figure \ref{fig:SAR-trajectory-and-signals}(c).

\begin{equation}
\Delta R(\tau)=d_{t}(\tau)-d_{c}(\tau)\label{eq:differential-range}
\end{equation}
Using differential range, the SAR phase history output in the receiver
from the arbitrary target location $p_{t}=(x,y,z)$ is given by Equation
(\ref{eqn: complex_phase_history_data})
\begin{equation}
S(f_{k},\tau_{n})=A(f_{k},\tau_{n})\exp\left(\frac{-j4\pi f_{k}\Delta R(\tau_{n})}{c}\right)\label{eqn: complex_phase_history_data}
\end{equation}
where, pulse index, $\tau_{n}=(\tau_{1},\tau_{N_{p}})$, and frequency
index, $f_{k}=(f_{1},f_{K})$. Unknown quantities that must be calculated
via geometric optics include the geometric slant range between target
scene locations and the antenna, $d_{t}(\tau)$ and the intensity
of the reflection $A(f_{k},\tau_{n})$. Determination of $d_{t}(\tau)$
allows $\Delta R(\tau)$ to be computed and, by extension, the differential
round-trip time of arrival between a target point reflection and a
phase center point reflection. Note that this time difference can
be made explicit via the substitution of this time difference as $\Delta t(\tau)=\frac{2\Delta R(\tau)}{c}$
in equation (\ref{eqn: complex_phase_history_data}) where $2\Delta R(\tau)$
is the round trip differential distance. This substitution gives rise
to the time-domain phase history signal shown in Equation (\ref{eqn: complex_phase_history_time_domain}).

\begin{equation}
S(f_{k},\tau_{n})=A(f_{k},\tau_{n})\exp\left(-j2\pi f_{k}\Delta t(\tau_{n})\right)\label{eqn: complex_phase_history_time_domain}
\end{equation}

The target surface locations required to compute $d_{t}(\tau)$ across
the region illuminated by the SAR antenna are determined by geometric
ray tracing and the material reflectance properties which are attributed to
each surface element to resolve the backscatter intensity $A(f_{k},\tau_{n})$.
The OptiX simulator densely samples these intersection locations over
the aperture of the antenna to compute the simulated SAR phase history.

\subsubsection{NVIDIA OptiX Implementation: SAR Pulse and Phase History Simulation}

OptiX is used to hardware-accelerate the calculation of 3D intersection
locations of rays of electromagnetic radiation emitted into the scene.
Figure \ref{fig:SimSAR OptiX-block-diagram} shows the OptiX SAR simulation
processing blocks for single pulse phase history generation. The functional
role of each block in Figure \ref{fig:SimSAR OptiX-block-diagram}
is described in the following list:
\begin{figure}[htp]
\noindent \centering{}\includegraphics[height=2.25in]{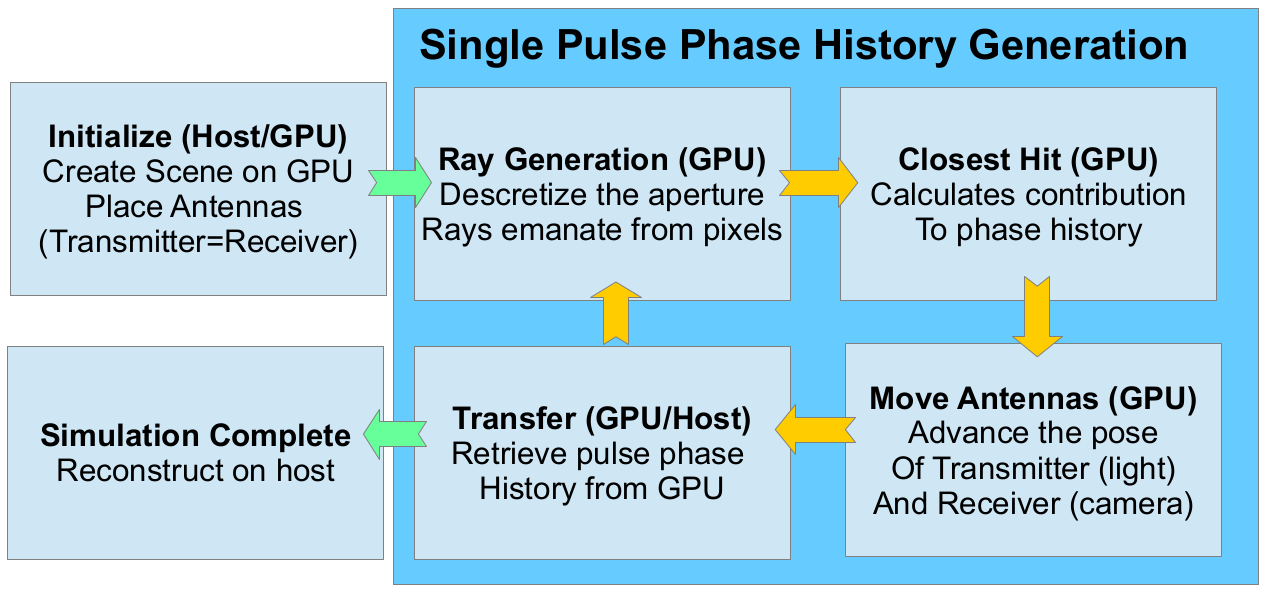}
\caption{\label{fig:SimSAR OptiX-block-diagram}Current SAR simulation software
using NVIDIA's OptiX ray tracing computational framework}
\end{figure}

\begin{enumerate}
\item \textbf{Initialize (Host \& GPU)}: All parameter files and data are
loaded from host storage and the 3D SAR simulation scene context is
created on the GPU device.
\item \textbf{Ray Generation (GPU)}: The ray generation program discretizes
the aperture and traces a random ray from each pixel in the descretize
antenna aperture. The OptiX framework constructs and maintains all
rays being traced.
\item \textbf{Closest Hit (GPU)}: The majority of the SAR pulse echo simulation
occurs within this program. OptiX invokes the ``closest hit'' program
for each ray that intersects a scene object. The scene intersection
location and intensity of the incident ray are used to calculate the
contribution of this target to the current pulse phase history equation
(\ref{eqn: complex_phase_history_data}).
\item \textbf{Move Antennas (GPU)}: Once all rays for an antenna position
have been traced, the antenna is moved to the next position in the
trajectory and, if applicable, the antenna pose is updated. A new
pulse phase history is simulated by returning to step (2) until all
pulses have be integrated into the phase history, $S(f_{k},\tau_{n})$.
\item \textbf{Transfer (GPU $\longrightarrow$ Host)}: The phase history
values are transferred off the GPU device and back to the CPU host
for storage and potential use in SAR image reconstruction algorithms.
\end{enumerate}
The joint impact of these programs allows on-device hardware accelerated
computation of simulated SAR EM pulse phase history. The developed
OptiX program code is compatible with many NVIDIA devices allowing
for simulation on most contemporary NVIDIA GPUs. Further, the code
can utilize multiple GPUs to further accelerate calculation.

The ray generation step propagates EM waves into the scene and ray
tracing continues as these rays reflect within the scence until all
rays have either: (1) missed a surface or (2) been terminated. Ray
termination occurs when the ray's number of surface reflections, i.e.,
bounce levels, exceeds the user settings or if the ray energy decays
below a threshold. Higher quality simulation may generate rays and
accumulate the sensed echo multiple times which is common practice
for ray tracing in computer graphics contexts.

Contributions to the phase history are computed within the ``closest
hit'' program. Specifically, the ``closest hit'' program first
uses the surface reflectance to determine the magnitude of energy
reflected in the direction of the antenna. For surfaces having significant
backscatter energy in the direction of the antenna, we compute the
sensed magnitude, $A(f_{k},\tau_{n})$, and relative slant range,
$\Delta R(\tau)$, from Equation (\ref{eqn: complex_phase_history_data})
and add the appropriate contributions to phase history for each frequency.
The location of the surface intersection provides $d_{t}(\tau)$ from
which $\Delta R(\tau)$ can be computed giving a phase value $\psi(f_{k},\tau_{n})=j4\pi f_{k}\frac{\Delta R(\tau_{n})}{c}$.
The magnitude, $A(f_{k},\tau_{n})$, is computed from the incident
ray intensity and the local surface reflectance and apparent size
of the antenna from the perspective of the surface (foreshortening).
Rays with significant reflected energy in directions other than the
antenna emit new rays in the reflected direction for ``multi-bounce''
ray simulation.

\subsubsection{OptiX Simulation with Single Precision Float Calculations\label{subsec:Numerical_precision}}

One shortcoming of simulation using the OptiX ray tracing library
is that \emph{all OptiX accelerated computations are restricted to
use single precision float values}. For SAR simulation this is a challenge
due to the relative scale of the emitted EM frequency wavelengths
and the distances over which these waves propagate in simulation to
calculate the phase history. 

Consider a simulation situation where an aerial vehicle at 10km slant
range emits a X-band pulse of RF energy as a SAR RF pulse. Given that
the wavelength of the emitted X-band frequencies are close to $3$cm,
there is a significant difference in the relative size of the numbers
in the phase calculation. More specifically, the phase calculation
of equation (\ref{eqn: complex_phase_history_data}) for frequency
$f_{k}$ is given by $\psi(f_{k},\tau_{n})=j4\pi f_{k}\frac{\Delta R(\tau_{n})}{c}$.
This calculation involves very large numbers, e.g., $c=0.299792458*10^{9}$
m/sec and $f_{k}$ having magnitude $10*10^{9}$, and much smaller
numbers, e.g., $\Delta R(\tau_{n})$ which has a maximum size of the
range swath width ($<10000$ m). Direct calculation of the ratios
involving large and small numbers, e.g., $\frac{\Delta R(\tau_{n})}{c}$,
leads to significant truncation error due to limitations in the representation
accuracy of single precision float numbers. We limit truncation error
by representing frequencies $f_{k}$ as frequency relative to the
center frequency, $f_{c}$, e.g., $f_{k}=f_{c}+k\Delta f$ where $k$
is an integer ($k=-\frac{K-1}{2},\ldots,\frac{K-1}{2}$ for $K$ frequencies)
and $\Delta f=\frac{B}{N_{f}-1}$ where $B$ denotes the pulse bandwidth.
Substituting this into the equation for phase we get $\psi(f_{k},\tau_{n})=j4\pi(f_{c}+k\Delta f)\frac{\Delta R(\tau_{n})}{c}$.
At this point we compute the constants ($\frac{f_{c}}{c}$, $\frac{\Delta f}{c}$)
using double precision and include these as constants in our single
precision phase calculation which is now factored as a constant term
$\psi(f_{k},\tau_{n})=j4\pi\frac{f_{c}}{c}+\frac{\Delta f}{c}k\Delta R(\tau_{n})$
limiting the on-device calculation to $k\Delta R(\tau_{n})$ which
can be multiplied by double precision values for ($\frac{f_{c}}{c}$,
$\frac{\Delta f}{c}$) on the host.

\section{Results}

Our experimental results are provided by analyzing the output of the
simulator along (3) distinct dimensions of performance:
\begin{enumerate}
\item Computational performance experiments evaluate the speed of the OptiX-accelerated
ray tracing for SAR simulation.
\item Numerical performance experiments evaluate the accuracy of the computed
phase history solution.
\item Reconstruction performance experiments evaluate the effect of numerical
errors from (2) on the output of two SAR image reconstruction algorithms.
\end{enumerate}
Analysis details trade-offs in terms of performance benefits
and accuracy costs associated with SAR simulation using OptiX accelerated
GPU programming. We use as a control the SAR simulator provided in
\cite{Gorham2010SARIF} which simulates SAR in MATLAB including computation
of the phase history for multiple point target reflectors and reconstruction
of 2D SAR images using the matched filter and backprojection algorithms.

\subsection{Computational Performance Results}

Our evaluation of SAR simulation performance compares the ray tracing
speed of the OptiX acclerated simulator against that of the MATLAB
simulator. We then analyze the relative speed of 4 different NVIDIA
GPUs ranging from inexpensive laptop GPUs (GeForce 940M), inexpensive
desktop GPUs (Quadro P620), more expensive GPUs (Quadro M5000) and
recently released NVIDIA RTX GPUs which perform both massively parallel
and hardware-accelerated ray tracing. We then investigate the benefit
of using multiple GPUs to distribute the computation of ray tracing
results.

Our first computational performance experiment benchmarks the MATLAB
vs. OptiX speed by duplicating the three point target SAR simulation
discussed in the results of Gorham et. al \cite{Gorham2010SARIF}.
Table \ref{tab:OptiX-vs-MATLAB} contains the performance results
found for this simulation. It is not surprising to speedups
in excess of 100,000 considering the MATLAB code is in an interpreted
language (Java) and does not benefit from any acceleration.

\begin{table}[hbt!]
\noindent \begin{centering}
\begin{tabular}{|c|c|c|c|}
\hline 
\textbf{Algorithm}  & \textbf{Device}  & \textbf{Rays traced/sec}  & \textbf{Speed-up factor}\tabularnewline
\hline 
\hline 
\textbf{Gorham et. al. \cite{Gorham2010SARIF}}  & CPU  & 156.38 rays/sec  & -\tabularnewline
\hline 
\textbf{SimSAR OptiX}  & RTX 2070 (GPU)  & \textbf{1.53 Mrays/sec}  & \textbf{100,000 x}\tabularnewline
\hline 
\end{tabular}
\par\end{centering}
\centering{}\caption{\label{tab:OptiX-vs-MATLAB}Three point target simulation MATLAB CPU
vs OptiX speed performance comparison.}
\end{table}

We then perform the same simulation of the three point target SAR
simulation on (4) different NVIDIA OptiX GPUs. Table \ref{tab:tabulated-ray-counts}
contains the number of primary, secondary and tertiary rays traced
during the SAR simulation of the experimental scene. Primary rays
are emitted as a SAR pulse and are generated by the ray generation
program. Secondary and tertiary rays are generated when surfaces reflect
emitted rays once or twice respectively. The tabulated results of
this experiment are shown in Table \ref{tab:tabulated-ray-counts}.
The most significant finding is an order of magnitude
($\sim$x10) increase in computational performance afforded by the
hardware accelerated RTX technology.

\begin{table}[hbt!]
\noindent \begin{centering}
\begin{tabular}{|c|c|c|c|c|}
\hline 
\textbf{Device}  & \begin{tabular}{@{}c@{}}\textbf{GeForce} \\ \textbf{940M}\end{tabular} &
\begin{tabular}{@{}c@{}}\textbf{Quadro} \\ \textbf{M5000}\end{tabular} &
\begin{tabular}{@{}c@{}}\textbf{Quadro} \\ \textbf{P620}\end{tabular} &
\begin{tabular}{@{}c@{}}\textbf{RTX} \\ \textbf{2070}\end{tabular}\tabularnewline
\hline 
\hline
\textbf{Number of frequencies}  & 512  & 512  & 512  & 512 \tabularnewline
\hline
\textbf{Number of pulses}  & 128  & 128  & 128  & 128 \tabularnewline
\hline
\textbf{Primary rays (Mrays)}  & 4.43  & 7.05  & 6.26  & 45.93 \tabularnewline
\hline 
\textbf{1 bounce rays (Mrays)}  & 4.45  & 8.16  & 7.37  & 54.71 \tabularnewline
\hline 
\textbf{2 bounce rays(Mrays)}  & 2.39  & 2.84  & 2.22  & 15.98 \tabularnewline
\hline 
\textbf{Closesthit (Mrays)}  & 3.03  & 5.60  & 1.67  & 61.95  \tabularnewline
\hline 
\textbf{Total traced rays (Mrays)}  & 14.30  & 24.04  & 17.53  & \textbf{178.57} \tabularnewline
\hline 
\textbf{Total Time (sec)}  & 141.462  & 140.794  & 146.403  & 116.572 \tabularnewline
\hline 
\textbf{Performance (Mrays/s)}  & 0.10  & 0.17  & 0.12  & 1.53 \tabularnewline
\hline 
\end{tabular}
\par\end{centering}
\centering{}\caption{\label{tab:tabulated-ray-counts}Contribution of closest hit program
and various depths of ray generation program in total ray counts. for
three point target.}
\end{table}

Experiment 3 constructs a scene from simplistic primitive shapes as
provided by the YAML ``primitive'' object and some sample 3D triangular
mesh objects. These objects and the number of vertices and triangles
required to specify the object geometries are contained in Table \ref{tab:model_polygon_count}.

\begin{table}[hbt!]
\noindent \begin{centering}
\begin{tabular}{|c|c|c|}
\hline 
\textbf{Primitive type}  & \textbf{Vertices}  & \textbf{Triangles}\tabularnewline
\hline 
\hline 
\textbf{Plane}  & 4  & 2 \tabularnewline
\hline 
\textbf{Box}  & 24  & 12 \tabularnewline
\hline 
\textbf{Sphere}  & 16290  & 32040 \tabularnewline
\hline 
\textbf{Torus}  & 32761  & 64800 \tabularnewline
\hline 
\textbf{T72.obj (tank in figure \ref{fig:YAMLObjects}(b))}  & 11348  & 18064 \tabularnewline
\hline 
\textbf{box.obj} & 24  & 24 \tabularnewline
\hline 
\end{tabular}
\par\end{centering}
\centering{}\caption{\label{tab:model_polygon_count}Vertex and Polygon counts for experimental
3D scene objects.}
\end{table}

Our third computational performance experiment constructs a 3D scene
from two tank mesh objects and inserts additional primitives into
the scene including a plane and a box primitive. This scene has $\sim$22k
(22724) vertices and $\sim$36k (36142) triangles as indicated by
Table \ref{tab:model_polygon_count}. It is important to note that
the computational difficulty of the ray tracing result depends strongly
on the number of polygons and (through the BVH hierarchy) the geometric
distribution of the polygons in terms of local clustering. For example,
scenes consisting of many polygons are more computationally intensive
than lower polygon count scenes and when these polygons are non-uniform
in size and spatial distribution, the computational costs for ray tracing
increase. Table \ref{tab:model_polygon_count} contains our observed
computational performance for ray tracing simulation on an RTX 2070
GPU. 
\begin{table}[tbh]
\noindent \begin{centering}
\begin{tabular}{|c|c|}
\hline 
\textbf{Target}  & \textbf{Scene 2 (36k triangles)} \tabularnewline
\hline 
\hline 
\textbf{Device}  & RTX 2070 \tabularnewline
\hline 
\textbf{Number of frequencies}  & 512 \tabularnewline
\hline 
\textbf{Number of pulses}  & 128 \tabularnewline
\hline 
\textbf{Number of Vertices}  & 22724 \tabularnewline
\hline 
\textbf{Number of Triangles}  & 36154 \tabularnewline
\hline 
\textbf{Primary rays (Mrays)}  & 176.96\tabularnewline
\hline 
\textbf{1 bounce rays (Mrays)}  & 184.97\tabularnewline
\hline 
\textbf{2 bounce rays(Mrays)}  & 108.15\tabularnewline
\hline 
\textbf{Closesthit (Mrays)}  & 162.39\tabularnewline
\hline 
\textbf{Total traced rays (Mrays)}  & \textbf{632.47}\tabularnewline
\hline 
\textbf{Total Time (sec)}  & 148.477\tabularnewline
\hline 
\textbf{Performance (Mrays/s)}  & 4.26\tabularnewline
\hline 
\end{tabular}
\par\end{centering}
\centering{}\caption{\label{tab:table 6}Performance of the OptiX simulator for the scene
shown in Figure \ref{fig:YAMLObjects}(b) which contains $\sim$36k
polygons. }
\end{table}

In summary, our results indicate that very large computational benefits
are gained through the use of the OptiX ray tracing framework for
simulation. Further, results show that, for our experimental scene,
NVIDIA RTX hardware-accelerated computation provides an additional
order of magnitude of compute acceleration beyond other GPU software
implementations of OptiX.

\subsection{Phase History Numerical Accuracy Results\label{subsec:Results-Phase-History-Numerical}}

Our evaluation of SAR simulation numerical performance compares the
values of the computed phase history of the OptiX accelerated simulator
against that of the MATLAB simulator. This is accomplished
by taking the difference between the complex numbers of the OptiX
and MATLAB phase histories. Let $S_{G}(f_{k},\tau_{n})$ and $S_{O}(f_{k},\tau_{n})$
denote $K\times N_{p}$ phase history images formed by a SAR simulation
using of $k=1,\ldots,K$ discrete range frequencies and $n=1,\ldots,N_{p}$
cross-range pulses. We then consider the average numerical errors
per frequency (for each range cell) as shown in Equation (\ref{eq:phase_history_range_avg_numerical_error})
and per pulse (for each cross-range measurement) as shown in Equation
(\ref{eq:phase_history_crossrange_avg_numerical_error}) respectively.
\begin{figure}
\noindent \centering{}\subfloat[Frequency (range cell) numerical error.]{\noindent \begin{centering}
\includegraphics[width=3in,height=2in]{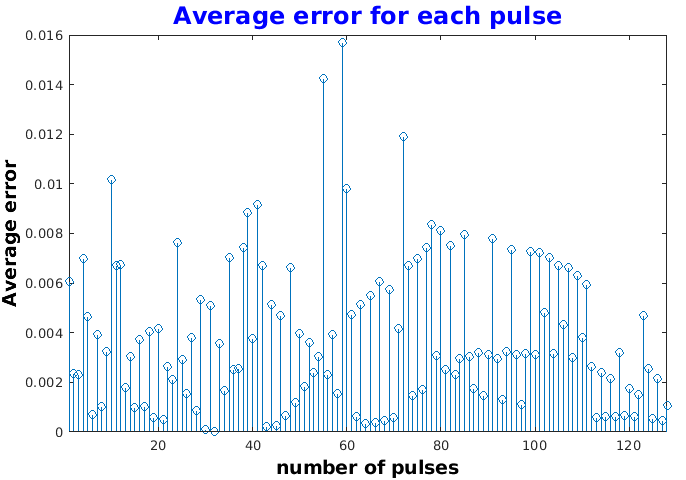} 
\par\end{centering}
}\subfloat[Pulse (cross-range) numerical error.]{\noindent \begin{centering}
\includegraphics[width=3in,height=2in]{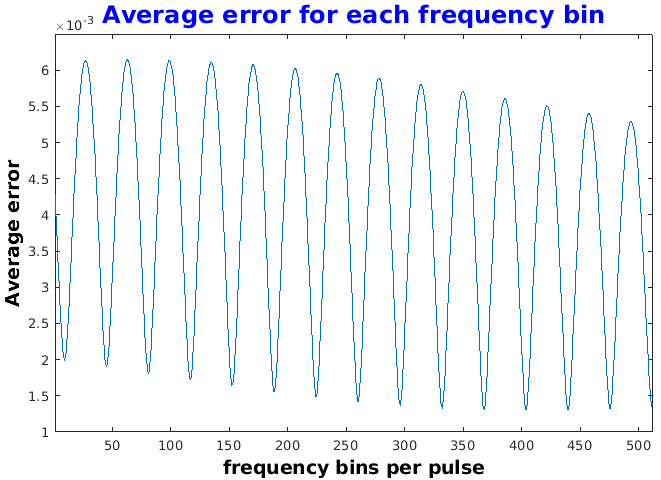} 
\par\end{centering}
}

\caption{\label{fig:phase_history_numerical_accuracy}(a,b) show numerical
error between single precision OptiX SAR phase histories and double
precision MATLAB phase histories. (a) plots the average numerical
range cell error for a 512 frequency SAR simulation, $\epsilon_{f}(f_{k})$.
(b) shows the average numerical cross-range cell error for a 128 pulse
SAR simulation.}
\end{figure}
\begin{equation}
\epsilon_{f}(f_{k})=\frac{1}{N_{p}}\sum_{n=1}^{N_{p}}\left|S_{O}(f_{k},\tau_{n})-S_{G}(f_{k},\tau_{n})\right|\label{eq:phase_history_range_avg_numerical_error}
\end{equation}

\begin{equation}
\epsilon_{\tau}(\tau_{n})=\frac{1}{K}\sum_{k=1}^{K}\left|S_{O}(f_{k},\tau_{n})-S_{G}(f_{k},\tau_{n})\right|\label{eq:phase_history_crossrange_avg_numerical_error}
\end{equation}

Results shown in Figure \ref{fig:phase_history_numerical_accuracy}
show numerical errors. The numerical error is controlled by reorganizing
the sensitive phase computation to re-express this calculation in
terms of similar magnitude values to make full use of all available
bits of precision in the single precision number format as described
in \S~\ref{subsec:Numerical_precision}. The problem stems from
the fact that single precision float number format has 24 precision
bits. Accuracy in floating point multiplication for the number pair,
$AB$, depends on the relative size of the two numbers. Specifically,
large differences in the floating point exponent limits the accuracy
of the calculated result. We maximize precision when these exponents
are identical and, under these circumstances, the largest number of
precision bits enter into the computed result. Identical exponents
occur when the two numbers being multiplied are within a factor of two.
The modification of \S~\ref{subsec:Numerical_precision} achieves
this goal by refactoring the phase calculation.

\subsection{SAR Image Reconstruction Accuracy Results}

Our evaluation of SAR simulation reconstruction performance compares
the values of SAR images reconstructed from simulated phase history
images provided by the OptiX acclerated simulator against that of
the MATLAB simulator. Our results compare reconstruction results using
two SAR image formation algorithms: (1) the matched filter algorithm
and (2) the backprojection algorithm.

\subsubsection{Matched Filtering and Backprojection Reconstruction Algorithms}

SAR image formation from matched filtering is an optimal method to
reconstruct the spatial $(x,y)$ reflectance of scene targets. It
uses optimal detection theory to extract the known frequency responses
present in the phase history and provides the maximum signal to noise
ratio (SNR) image reconstruction assuming the phase history is perturbed
by Gaussian noise. Matched filter reconstruction tailors a frequency-specific
filter that will ideally match to the target. The matched filter response
for target $p_{t}$ can be calculated using Equation (\ref{eq:matched_filter})
\cite{Gorham2010SARIF}, 
\begin{equation}
I_{mf}(p_{t})=\frac{1}{N_{p}K}\sum_{n=1}^{N_{p}}\sum_{k=1}^{K}S(f_{k},\tau_{n})\exp\left(\frac{j4\pi f_{k}\Delta R(\tau_{n})}{c}\right)=A_{0}\label{eq:matched_filter}
\end{equation}

SAR image formation from the backprojection algorithm uses the theory
of the inverse Radon transform to reconstruct the spatial $(x,y)$
reflectance of scene targets. Equation (\ref{eq:bp2}) details the
computation required to calculate the backprojection response for
target $p_{t}$.
\begin{equation}
I(p_{t})=\sum_{n=1}^{N_{p}}s_{int}(p_{t},\tau_{n})\label{eq:bp2}
\end{equation}
where, $s_{int}(p_{t},\tau_{n})$ are values that are interpolated
from the range profile, $s(m,\tau_{n})$. Equation (\ref{eqn: bp range profile at range bin m})
is the ``range profile'' value taken from the phase history at range
bin $m$ and pulse $\tau_{n}$ , collected by $N_{p}$ pulses over
a range of $K$ frequencies, is 
\begin{equation}
s(m,\tau_{n})=\sum_{k=1}^{K}S(f_{k},\tau_{n})\exp\left(\frac{j4\pi f_{k}\Delta R(m,\tau_{n})}{c}\right)\label{eqn: bp range profile at range bin m}
\end{equation}
After substituting $f_{k}=(k-1)\Delta f+f_{0}$ in the above equation
and performing some algebric manipulation, Equation (\ref{eqn: bp range profile at range bin m})
can be expressed in terms of the inverse discrete Fourier transform
as shown in Equation (\ref{eq:bp}) \cite{Gorham2010SARIF}.
\begin{figure}
\noindent \begin{centering}
\subfloat[MATLAB]{\noindent \begin{centering}
\includegraphics[height=1.4in]{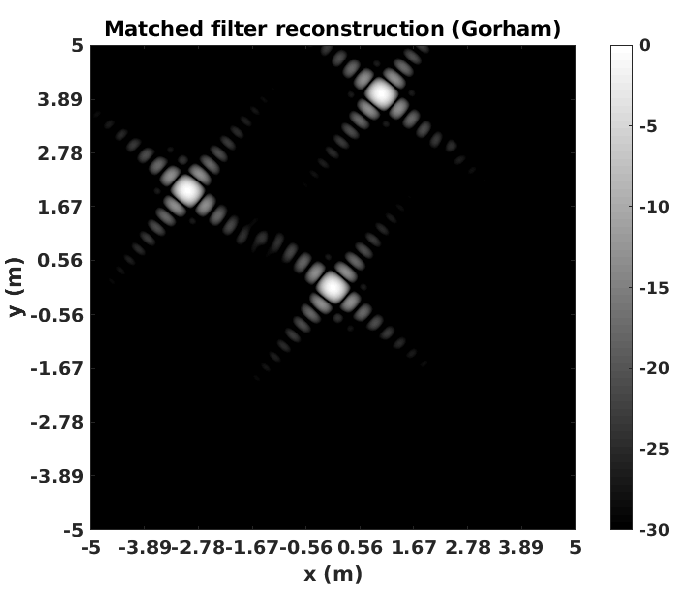} 
\par\end{centering}
}\subfloat[OptiX simulator]{\noindent \begin{centering}
\includegraphics[height=1.4in]{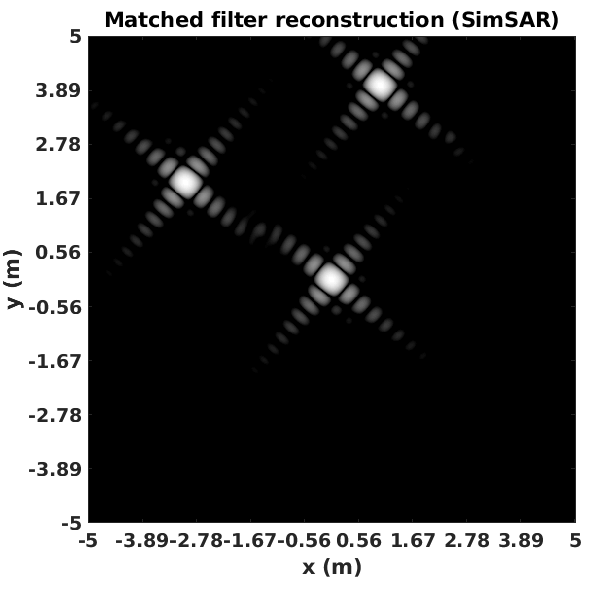} 
\par\end{centering}
}\subfloat[MATLAB]{\noindent \begin{centering}
\includegraphics[height=1.4in]{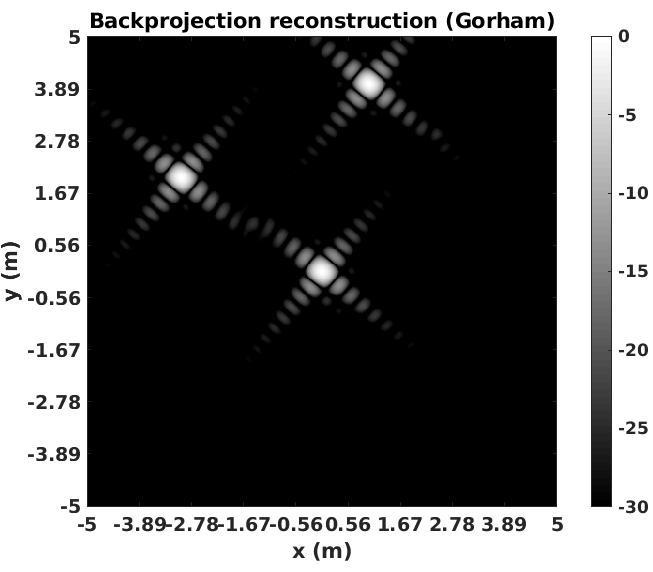} 
\par\end{centering}
}\subfloat[OptiX simulator]{\noindent \begin{centering}
\includegraphics[height=1.4in]{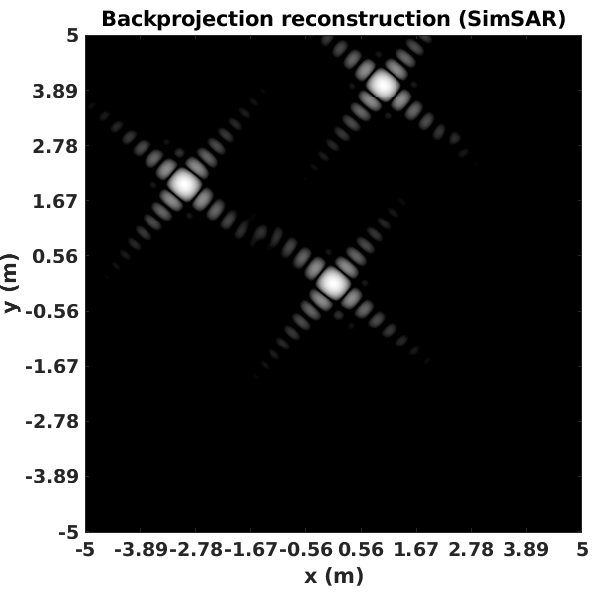} 
\par\end{centering}
}
\par\end{centering}
\caption{\label{fig:SAR reconstruction}(a,b) show MATLAB\cite{Gorham2010SARIF}
and OptiX simulator SAR image reconstructions for 3 point targets
using the matched filter algorithm. (c,d) show MATLAB and OptiX simulator
SAR image reconstructions for 3 point targets using back projection
algorithm.}
\end{figure}

\begin{equation}
s(m,\tau_{n})=N_{p}\cdot\{FT^{-1}(S(f_{k},\tau_{n}))\}\exp\left(\frac{j2\pi f_{1}(m-1)}{N_{p}\Delta f}\right)\label{eq:bp}
\end{equation}

Computation of the backprojection image is less computationally costly
than matched filtering with a computational complexity of $O(N^{3})$.
The backprojection algorithm can also take advantage of parallel
processing and can be implemented on graphics processing units (GPU)
\cite{bpGPU1,bpGPU2}.
\begin{figure}
\noindent \centering{}\includegraphics[width=10cm]{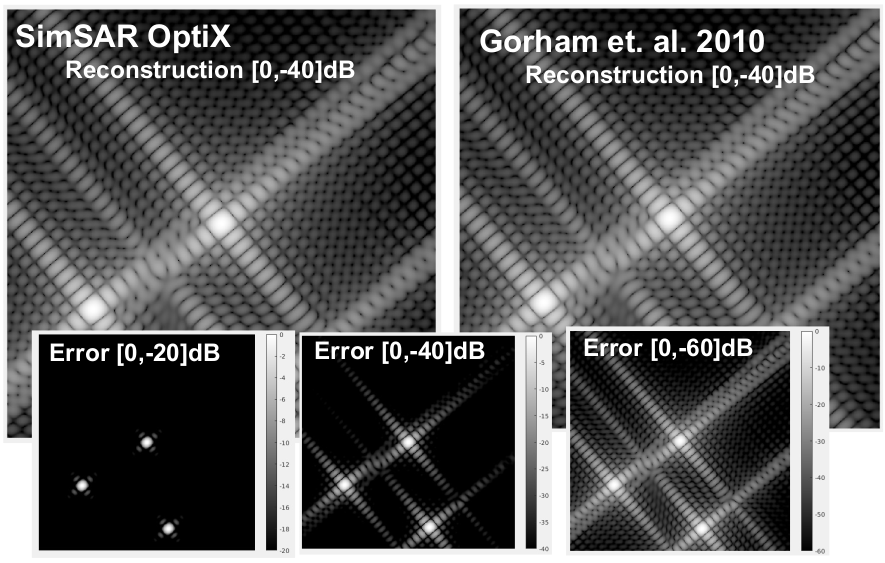}
\caption{\label{fig:SimSAR_result_compare}Numerical validation by forming
a SAR image of a 3-point test target located having 3D locations (0,0,0),
(-3,2,0), and (1,4,0). (left) shows OptiX accelerated simulation results.
(right) shows results from Gorham et. a l. \cite{Gorham2010SARIF}.
Bottom figures show three images highlighting small ($<$0.1 percent)
numerical differences in the reconstructed SAR image values.}
\end{figure}

\subsubsection{Reconstruction accuracy results}

Figure \ref{fig:SAR reconstruction} (a-d) shows SAR images formed
for the 3 point target simulation duplicated from Gorham et. al. Phase
history data was simulated for 3 point targets using Equation \ref{eqn: complex_phase_history_data}
where $A(f_{k},\tau_{n})$ = 1. Similar to Gorham et. al., $N_{p}$
= 128 pulses were simulated with $K$ = 512 frequency samples per
pulse, a center frequency of 10 GHz and a 600 MHz bandwidth. A circular
flight path was used with a 30 degree depression angle and a slant
range of 10 km. A 3 degree integration angle was used with a center
azimuth angle of 50 degrees. The scene extent was 10 m x 10 m with
2 cm pixel spacing in each dimension. We defined all the above parameters
in an YAML file.

Figure \ref{fig:SAR reconstruction}(a,b) shows matched filter reconstructions
and Figure \ref{fig:SAR reconstruction}(c,d) shows backprojection
algorithm reconstructions for the OptiX simulator and the MATLAB simulator
from Gorham et. al. While the images shown in this figure look very
similar, one would suspect they are numerically distinct given the
results of \S~\ref{subsec:Results-Phase-History-Numerical}.

Finally, Figure \ref{fig:SimSAR_result_compare} shows both the reconstructed
SAR images and the differences between the reconstructed images. Under
this analysis we are able to see that, as one would suspect, larger
errors in reconstruction occur in the vicinity of high-reflectance
spatial locations.

\section{Conclusion}

This article proposes a Shooting and Bouncing Ray (SBR) approach for
SAR simulation using NVIDIA's OptiX library to accelerate the computationally
expensive ray tracing calculations required by this approach. The
computational performance of this approach provides orders of magnitude
speed increases over CPU simulation and even more when using RTX graphics
cards for simulation which can further accelerate the GPU calculations
by a factor of 10 using ray tracing specific hardware. The new structure,
processing framework and calculations required by the OptiX ray tracing
library are described that allow the simulation software to run using
one or more massively parallel GPU computation devices to obtain solutions
to the geometric optics simulation at unprecedented speeds. The shortcoming
of the OptiX library's restriction to single precision float representation
is discussed and modifications of sensitive calculations are proposed
to reduce truncation error thereby increasing the simulation accuracy
under this constraint. Computational performance is validated against
one of the few open source SAR simulators available and demonstrate
massive speed increases. Analysis is also made on the errors associated
with simulation using single precision floats versus double precision
floats and the impact of this restriction in accuracy is analyzed
in both the calculation of the simulated phase history and for the
reconstruction of the SAR spatial reconstruction images. 

 \bibliographystyle{spiebib}
\bibliography{2020_SPIE_SimSAR_OptiX}

\begin{thebibliography}{10}

\bibitem{radartutorial}
C.~Wolff, {\em https://www.radartutorial.eu/20.airborne/ab07.en.html}, November
  1999.

\bibitem{TargetRecognition}
S.~{Chen} and H.~{Wang}, ``{SAR} target recognition based on deep learning,''
  in {\em 2014 International Conference on Data Science and Advanced Analytics
  (DSAA)},  pp.~541--547, 2014.

\bibitem{Mapping}
J.~{Zhang}, S.~{Yang}, Z.~{Zhao}, and G.~{Huang}, ``{SAR} mapping technology
  and its application in difficulty terrain area,'' in {\em 2010 IEEE
  International Geoscience and Remote Sensing Symposium},  pp.~3608--3611,
  2010.

\bibitem{Ocean_Surveillance}
M.~Yeremy, J.~Campbell, K.~Mattar, and T.~Potter, ``Ocean surveillance with
  polarimetric sar,'' {\em Canadian Journal of Remote Sensing}~{\bf 27}(4),
  pp.~328--344, 2001.

\bibitem{Geology}
Z.~Perski, ``Application of {SAR} imagery and {SAR} interferometry in digital
  geological cartography,'' in {\em The Current Role of Geological Mapping in
  Geosciences},  S.~R. Ostaficzuk, ed., pp.~225--244, Springer Netherlands,
  (Dordrecht), 2005.

\bibitem{ForestMapping}
A.~Pulella, R.~Aragão~Santos, F.~Sica, P.~Posovszky, and P.~Rizzoli,
  ``Multi-temporal sentinel-1 backscatter and coherence for rainforest
  mapping,'' {\em Remote Sensing}~{\bf 12}(5), 2020.

\bibitem{mitocwSAR}
G.~Charvat, J.~Williams, A.~Fenn, S.~Kogon, and J.~Herd, {\em RES.LL-003 Build
  a Small Radar System Capable of Sensing Range, Doppler, and Synthetic
  Aperture Radar Imaging}, January 2011.

\bibitem{Carrara1995}
W.~Carrara, {\em Spotlight synthetic aperture radar : signal processing
  algorithms}, Artech House, Boston, 1995.

\bibitem{2015arXivHowMuchData}
J.~{Cho}, K.~{Lee}, E.~{Shin}, G.~{Choy}, and S.~{Do}, ``{How much data is
  needed to train a medical image deep learning system to achieve necessary
  high accuracy?},'' Nov. 2015.

\bibitem{Tremblay_2018_CVPR_Workshops}
J.~Tremblay, A.~Prakash, D.~Acuna, M.~Brophy, V.~Jampani, C.~Anil, T.~To,
  E.~Cameracci, S.~Boochoon, and S.~Birchfield, ``Training deep networks with
  synthetic data: Bridging the reality gap by domain randomization,'' in {\em
  The IEEE Conference on Computer Vision and Pattern Recognition (CVPR)
  Workshops},  June 2018.

\bibitem{AnderssonNSSA19}
P.~Andersson, J.~Nilsson, M.~Salvi, J.~B. Spjut, and T.~Akenine{-}M{\"{o}}ller,
  ``{T}emporally {D}ense {R}ay {T}racing,'' in {\em High-Performance Graphics
  2019 - Short Papers, Strasbourg, France, July 8-10, 2019},  pp.~33--38, 2019.

\bibitem{SBR_Lee}
H.~{Ling}, R.~. {Chou}, and S.~. {Lee}, ``Shooting and bouncing rays:
  calculating the {RCS} of an arbitrarily shaped cavity,'' {\em IEEE
  Transactions on Antennas and Propagation}~{\bf 37}(2), pp.~194--205, 1989.

\bibitem{RayTube_Lee}
S.~W. Lee, H.~Ling, and R.~Chou, ``Ray-tube integration in shooting and
  bouncing ray method,'' {\em Microwave and Optical Technology Letters}~{\bf
  1}(8), pp.~286--289, 1988.

\bibitem{Andersh2000Xpatch4T}
D.~J. Andersh, J.~Moore, S.~Kosanovich, D.~B. Kapp, R.~Bhalla, R.~Kipp,
  T.~Courtney, A.~Nolan, F.~German, J.~Cook, and J.~Hughes, ``Xpatch 4: the
  next generation in high frequency electromagnetic modeling and simulation
  software,'' 2000.

\bibitem{OSV_Radar}
{JRM Technologies}, {\em OSV Radar}, 2019.

\bibitem{Raytheon}
{Raytheon Corporation}, ``Raytheon inc., technology today: Highlighting
  raytheon’s technology,'' pp.~34--35, 2013.

\bibitem{StephenAuerPhDthesis}
S.~J. Auer, {\em 3D Synthetic Aperture Radar Simulation for Interpreting
  Complex Urban Reflection Scenarios}.
\newblock Dissertation, Technische Universität München, München, 2011.

\bibitem{Weijie2014SARIS}
X.~Wei-jie, L.~W. Hua, W.~P. Fei, L.~Hai-lin, and Z.~Jin-dong, ``Sar image
  simulation for urban structures based on sbr,'' 2014.

\bibitem{RaySAR3D}
S.~{Auer}, R.~{Bamler}, and P.~{Reinartz}, ``{RaySAR - 3D SAR} simulator: Now
  open source,'' in {\em 2016 IEEE International Geoscience and Remote Sensing
  Symposium (IGARSS)},  pp.~6730--6733, July 2016.

\bibitem{alma991010725391504091}
S.~Auer, S.~Hinz, and R.~Bamler, ``Ray-tracing simulation techniques for
  understanding high-resolution sar images,'' {\em IEEE transactions on
  geoscience and remote sensing a publication of the IEEE Geoscience and Remote
  Sensing Society.}~{\bf 48}(3), pp.~1445,1456, 2010-03.

\bibitem{NvidiaRTXdev}
{NVIDIA Corporation}, {\em {NVIDIA RTX platform}}, 2018.

\bibitem{NVIDIA_OptiX_6}
{NVIDIA Corporation}, {\em {NVIDIA OptiX Ray Tracing Engine}}, 2018.

\bibitem{Parker_optix:a}
S.~G. Parker, J.~Bigler, A.~Dietrich, H.~Friedrich, J.~Hoberock, D.~Luebke,
  D.~McAllister, M.~McGuire, K.~Morley, A.~Robison, and M.~Stich, ``Optix: A
  general purpose ray tracing engine,'' {\em ACM Trans. Graph.}~{\bf 29}, July
  2010.

\bibitem{BVH}
Y.~Gu, Y.~He, K.~Fatahalian, and G.~Blelloch, ``Efficient {BVH} construction
  via approximate agglomerative clustering,'' in {\em Proceedings of the 5th
  High-Performance Graphics Conference},  {\em HPG ’13}, p.~81–88,
  Association for Computing Machinery, (New York, NY, USA), 2013.

\bibitem{YAML}
O.~Ben-Kiki, C.~Evans, and I.~d{\"o}t Net, ``{YAML} ain’t markup language
  ({YAML™}) version 1.2,'' 2001.

\bibitem{McGuire2009Photon}
M.~McGuire and D.~Luebke, ``Hardware-accelerated global illumination by image
  space photon mapping,'' in {\em {ACM} {SIGGRAPH}/{EuroGraphics} High
  Performance Graphics 2009},  {ACM}, (New York, NY, USA), August 2009.
\newblock Proceedings of the 2009 ACM SIGGRAPH/EuroGraphics conference on High
  Performance Graphics.

\bibitem{Gorham2010SARIF}
L.~A. Gorham and L.~J. Moore, ``{SAR} image formation toolbox for {MATLAB},''
  in {\em Defense + Commercial Sensing},  2010.

\bibitem{bpGPU1}
T.~Hartley, A.~Fasih, C.~Berdanier, F.~Ozguner, and U.~Catalyurek,
  ``Investigating the use of {GPU}-accelerated nodes for {SAR} image
  formation,'' pp.~1 -- 8, 10 2009.

\bibitem{bpGPU2}
A.~Rogan and R.~Carande, ``{Improving the fast back projection algorithm
  through massive parallelizations},'' in {\em Radar Sensor Technology XIV},
  K.~I. Ranney and A.~W. Doerry, eds.,  {\bf 7669}, pp.~144 -- 151,
  International Society for Optics and Photonics, SPIE, 2010.

\end{thebibliography}

\end{document}